\documentclass[reprint,letterpaper,superscriptaddress,prl,aps,floatfix]{revtex4-2}
\usepackage{babel}
\usepackage{graphicx,mathptmx}
\usepackage{dcolumn}% Align table columns on decimal point
\usepackage{bm,bbm}% bold math
\usepackage{orcidlink,hyperref}
\usepackage{ulem}
\hypersetup{colorlinks = true,linkcolor = blue,anchorcolor = blue,citecolor = blue,filecolor = blue,urlcolor = blue}
\usepackage{enumerate,appendix}
\usepackage{amsmath,amsthm,amssymb,commath}
\usepackage{calc}
\usepackage{latexsym}

\newcommand{\ket}[1]{|#1\rangle}
\newcommand{\bra}[1]{\langle#1|}

\newcommand{\tr}{{\rm tr}}

\usepackage[charter,cal=cmcal,sfscaled=false]{mathdesign}
\usepackage{booktabs}
\usepackage{multirow}
\usepackage{subfigure}
\usepackage{dcolumn}
\usepackage{mathrsfs}
\usepackage{diagbox}
 \usepackage{array}
 \usepackage{makecell}

\newcommand{\dyad}[2]{\ensuremath{|\,{#1}\,\rangle\langle\,{#2}\,|}}
\usepackage{braket}

\begin{document}

\title{Efficient Experimental Qudit State Estimation via Point Tomography}

\author{D.~Mart\'inez\orcidlink{0000-0002-8154-7141}}
\affiliation{Departamento de F\'isica and Millennium Institute for Research in Optics, Universidad de Concepci\'on, 160-C Concepci\'on, Chile}
\affiliation{Vienna Center for Quantum Science and Technology (VCQ) and Christian Doppler Laboratory for Photonic Quantum Computer, Faculty of Physics, University of Vienna, 1090 Vienna, Austria}

\author{L.~Pereira\orcidlink{0000-0003-1183-2382}}
\affiliation{ICFO - Institut de Ciencies Fotoniques, The Barcelona Institute of Science and Technology, 08860 Castelldefels, Barcelona, Spain}

\author{K.~Sawada\orcidlink{0009-0003-2396-681X}}
\affiliation{Departamento de F\'isica and Millennium Institute for Research in Optics, Universidad de Concepci\'on, 160-C Concepci\'on, Chile}

\author{P.~Gonz\'alez}
\affiliation{Departamento de F\'isica and Millennium Institute for Research in Optics, Universidad de Concepci\'on, 160-C Concepci\'on, Chile}

\author{J.~Cari\~{n}e\orcidlink{0000-0001-6232-6012}}
\affiliation{Departamento de Ingenier\'ia El\'ectrica, Universidad Cat\'olica de la Sant\'isima Concepci\'on, Concepci\'on, Chile}

\author{M.~Muñoz}
\affiliation{Funda\c{c}\~{a}o Getulio Vargas, Rio de Janeiro, Brazil}

\author{A.~Delgado\orcidlink{0000-0002-8968-5733}}
\affiliation{Departamento de F\'isica and Millennium Institute for Research in Optics, Universidad de Concepci\'on, 160-C Concepci\'on, Chile}

\author{E.~S.~Gómez\orcidlink{0000-0003-3227-9432}}
\affiliation{Departamento de F\'isica and Millennium Institute for Research in Optics, Universidad de Concepci\'on, 160-C Concepci\'on, Chile}

\author{S.~P.~Walborn\orcidlink{0000-0002-3346-8625}}
\affiliation{Departamento de F\'isica and Millennium Institute for Research in Optics, Universidad de Concepci\'on, 160-C Concepci\'on, Chile}

\author{G.~Lima\orcidlink{0000-0001-7670-6032}}
\email{glima@udec.cl}
\affiliation{Departamento de F\'isica and Millennium Institute for Research in Optics, Universidad de Concepci\'on, 160-C Concepci\'on, Chile}

\date{\today}% It is always \today, today,

\begin{abstract}
Point tomography is a new approach to the problem of state estimation, which is arguably the most efficient and simple method for modern high-precision quantum information experiments. In this scenario, the experimenter knows the target state that their device should prepare, except that intrinsic systematic errors will create small discrepancies in the state actually produced. By introducing a new kind of informationally complete measurement, dubbed Fisher-symmetric measurements, point tomography determines deviations from the expected state with optimal efficiency. In this method, the number of outcomes of a measurement saturating the Gill-Massar limit for reconstructing a $d$-dimensional quantum states can be reduced from $\sim 4d-3$ to only $2d-1$ outcomes. Thus, providing better scalability as the dimension increases. Here we demonstrate the experimental viability of point tomography. Using a modern photonic platform constructed with state-of-the-art multicore optical fiber technology,  we  generate 4-dimensional quantum states and implement seven-outcome Fisher-symmetric measurements. Our experimental results exhibit the main feature of point tomography, namely a precision close to the Gill-Massar limit with a single few-outcome measurement. Specifically, we achieved a precision of $3.8/N$ while the Gill-Massar limit for $d=4$ is $3/N$ ($N$ being the ensemble size).
\end{abstract}

\maketitle

\paragraph{\textit{Introduction}--}High-dimensional quantum states (qudits) exhibit important advantages over bi-dimensional systems for quantum information processing. For instance, qudits can increase sensitivity in quantum metrology \cite{Lloyd,datta,albarelli2020} and efficiency in quantum computing \cite{Lanyon,Bocharov,Babazadeh,Muralidharan,Taddei_2020}. Nonetheless, the benefits of adopting qudit states may be overshadowed by the difficulty in estimating them accurately, impairing our capability to generate, control, and transmit them. This has led to the design of various estimation methods aimed at achieving high estimation precision or reducing the resources required by the estimation process. Modern adaptive quantum tomography methods, for example, rely on interactive algorithms to reach the Gill-Massar limit \cite{Gill} for some particular quantum states \cite{Ferrie,Mahler, Straupe2016, Struchalin2018, Huszar2012}. Unfortunately, in general, their efficiencies quickly degrade or the protocol become too complex to implement as the dimension of the system increases \cite{Pereira1}.

Single-setting tomographic methods provide an alternative to adaptive methods \cite{Renes}. In this approach, a single positive-operator-valued measure (POVM) is used for the state reconstruction. If the recorded statistics of the outcomes of the POVM are sufficient to reconstruct the quantum state, the POVM is said to be informationally complete. The POVM is also said to be globally informationally complete if it can be used to reconstruct any unknown quantum state, in which case the POVM must have at least $d^2$ elements \cite{Renes,Li} for a $d$-dimensional system. There have been several theoretical and experimental studies adopting globally informationally complete POVMs \cite{Appleby,MinimalTomography,MinimalTomographyExperimental,Bent15,stricker22}. Nevertheless, since the scalability of the number of outcomes is demanding, many of the experimental demonstrations actually do not fully implement the required generalized measurement. Instead, they rely on measuring the statistics of each outcome independently and, therefore, the simplicity provided by single-setting tomography is replaced by $d^2$ different measurements.

A modern take on POVM-based state reconstruction is provided by point tomography \cite{Li,Zhu}, which finds practical relevance for modern quantum platforms that can achieve high-precision in quantum information tasks. Point tomography departs from previous methods by relying not on globally- but instead on locally-complete POVMs. The new kind of measurement used, namely Fisher-symmetric measurements, are chosen such that the Fisher information is distributed uniformly among a set of parameters that uniquely identify quantum states in the neighborhood of an arbitrary target state. Thus, point tomography neatly applies to the situation where the experimentalist has a well-characterized preparation device that emits a state with only small systemic deviations from the target state. Choosing the proper POVM, the method can saturate the Gill-Massar limit.  Besides high-precision, point tomography has another huge practical advantage. By resorting only to locally complete POVMs, the number of POVM outcomes can be drastically reduced, giving much better scalability when applied to higher dimensions. In the case of pure quantum states, the number of outcomes can be reduced from $\sim 4d-3$ to only $2d-1$ \cite{Li,Zhu}. 

Here, we use state-of-the-art multicore optical fiber technology to demonstrate for the first time the experimental viability of this point tomography method. With this new photonic platform, we are able to generate path-encoded four-dimensional quantum states with a high degree of precision \cite{Carine}, and implement high-fidelity genuine seven-outcome POVMs \cite{Martinez2023}. Our experimental results exhibit the main feature of point tomography, namely a precision close to the Gill-Massar limit with a single few-outcome Fisher-symmetric measurement. A fit of the experimental data provides an estimation accuracy of $3.8/N$ while the Gill-Massar limit for $d=4$ is $3/N$ ($N$ being the ensemble size). This result is even valid for an ensemble as small as $N=50$. Furthermore, we also experimentally test the method for states in a broader neighborhood. In this case the estimation precision decreases, as expected, but we see that for small ensembles it is still possible to achieve a precision comparable to the Gill-Massar limit. Our results help pave the way for a broader adoption of point tomography, which resonates well with all modern quantum platforms being developed for high-precision quantum information processing \cite{ReviewQKD,ReviewQComputing}.

\paragraph{\textit{Method}--} As discussed above, a priori information is used in the construction of Fisher-symmetric measurements. That is, the state $|\psi\rangle$ to be estimated must be in the neighborhood of a given arbitrary fiducial state $|0\rangle$ 
\begin{equation}
    |\psi\rangle=\frac{1}{A}\left(|0\rangle+\sqrt{\theta}\sum_{j=1}^{3}|j\rangle\right), \label{state}
\end{equation}
where $A$ is a normalization constant, $\{|j\rangle\}$ with $j=1,\dots,d-1$ is an orthornormal basis, and the coefficient $\theta$ is infinitesimal $|\theta|^2\lll1$. The quality of the characterization depends on how small this parameter is. Then, the aim of point tomography is to estimate $|\psi\rangle$ with the ultimate precision provided by the Gill-Massar bound, while also maintaining the classical fisher information matrix uniformly distributed over the target state $|0\rangle$. As has been demonstrated in  Ref.~\cite{Li}, all rank-1 POVMs $|\phi^\eta\rangle\langle\phi^\eta|$ ($\eta=1,\dots,2d-1$), with $|\phi^\eta\rangle=\sum_{j=0}^{d-1} a^\eta_j|j\rangle$ and $a^\eta_0$ real, define a Fisher-symmetric measurement if the matrix $C$, with elements  $C_{j,k}=\sum_{\eta=1}^{2d-1}a_j^\eta a_k^\eta$ ($j,k=1,\dots,d-1$), has null norm (See supplemental material for details \cite{SuppMat}). 

Recently, a method to implement high-quality POVMs in a $d$-dimensional Hilbert space using $D \times D$ modern multiport beam splitters (MBS) was presented \cite{Martinez2023}. For a photonic qudit, encoded in terms of $d$ spatial optical modes, a rank-1 POVM is realized by connecting these modes to $d$ of the $D$ inputs of the MBS ($D > d$). Thus, by connecting different sets of inputs, there are ${D!}/[{d!(D-d)!}]$ different classes of possible POVMs that can be implemented. Exploiting the fact that relative phases can be imprinted between the optical modes before the MBS, the rank-1 POVM elements are proportional to the states
\begin{align}
\ket{\eta_j}=\Phi^\dagger_{k_1\cdots k_{d}} M_{k_1\cdots k_{d}} \ket{j}, \label{POVM}
\end{align}
where $k_1\cdots k_{d}$ defines the connected input modes of the MBS, $\Phi_{k_1\cdots k_{d}}$ is a diagonal matrix defining the phases applied before the MBS, and $M_{k_1\cdots k_{d}}$ is a $D \times d$ matrix, which is the part of the MBS  $D \times D$ unitary matrix acting on the modes $k_1\cdots k_{d}$ \cite{Martinez2023}. 

In this work, we are interested on implementing a fisher symmetric 7-outcome measurement onto a 4-dimensional fiducial state using the technique of Ref.~\cite{Martinez2023}. In this case, there are 35 different types of POVMs defined by Eq.~(\ref{POVM}) that could possibly be used in point tomography. For our particular implementation, these measurements are given explicitly in the supplemental material \cite{SuppMat}. In order to select the best POVM candidate, we minimize the 2-norm of the matrix $C$ for each family of feasible POVMs. We find that even though we cannot implement a Fisher symmetric measurement exactly, there is a configuration close to a perfect one, where the corresponding matrix $C$ has norm $|| C || \approx 0.63 $ (see supplemental material \cite{SuppMat}). As we demonstrate experimentally next, even in this case point tomography method works quite well.     

\begin{figure*}[t]
	\centering
    	\includegraphics[width=0.95\textwidth]{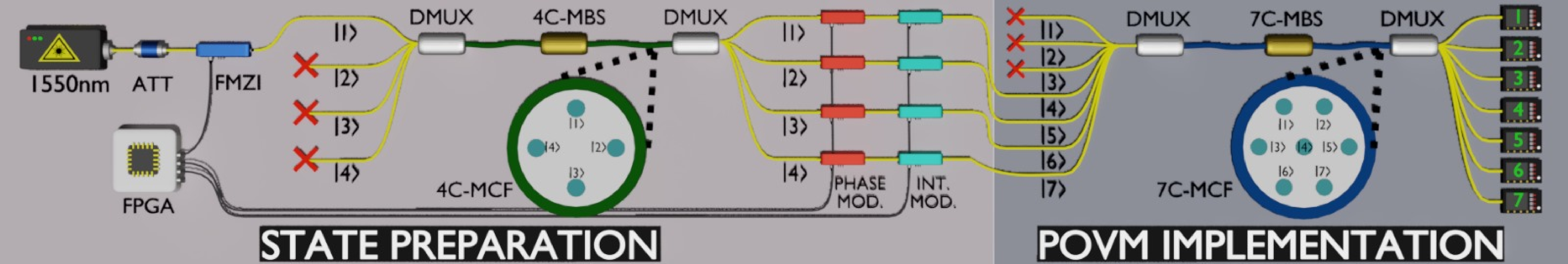}
    	\caption{Experimental setup. In the preparation stage, single photon states are generated with a CW-laser, an attenuator (Att), and an Mach-Zehnder based intensity modulator (FMZI). The light is distributed from one of the fiber cores to all four of them by a 4C-MBS (4-core multicore beam splitter). The cores are separated into fibers and the intensity and phase of each one are modulated by phase modulators (PMs) and variable intensity modulators (IM) to prepare four-dimensional single-photon states. In the measurement stage, the four fibers are fed into a 7C-MBS (7-core multicore beam splitter) and each of the seven outputs is sent to a single-photon detector (D1 through D7). Since the number of outcomes from the 7C-MBS is greater than 4, the corresponding measurement is a POVM. See the main text for details.}
    	\label{Figure1}
\end{figure*}

\paragraph{\textit{Experiment}--} Our experimental setup is depicted in Fig.~\ref{Figure1}. It has two main stages dedicated for state preparation and POVM implementation. The experiment relies on modern multi-core optical fiber (MCF) technologies, which have been developed to meet the increasing demand for bandwidth in optical communication networks \cite{Richardson_natphoton_2013}. In MCFs, light is transmitted through multiple single-mode cores that are contained within a common cladding. Since the crosstalk between them is depreciable, such core modes can be used to encode $d$-dimensional photonic quantum systems. Because the cores are within the same cladding, the platform is inherent robust against thermal and mechanical perturbations to the quantum system, as they act like global effects. In a series of independent studies (See Ref.~\cite{GuixReview_2019} for a comprehensive review), such platform has been shown to achieve high-precision in many different quantum information tasks, ranging from high-dimensional quantum communication \cite{YDing,Canas,Sarmiento,Bacco_2024}, entanglement generation \cite{Lee_2017, Lee_2019, Ortega,Gomez}, to more efficient quantum computing approaches \cite{Taddei_2020}. 

The purpose of the preparation stage is to generate $4$-dimensional quantum states. This is accomplished by using the path-encoding strategy for a single photon propagating over a 4-core fiber \cite{Carine}. The basis states $|i\rangle$, with $i=0,1,2,3$, denote the state of a photon transmitted by the $i$-th core of the MCF. In our setup, a semiconductor telecom CW-laser operating at 1546 nm, followed by an attenuator (Att), and an external fiber-pigtailed intensity modulator (IM), are used to generate weak coherent states comprised of 5 ns-long pulses at a rate of 2 MHz. The IM is controlled by a field programmable gate array unit (FPGA). The mean photon number per pulse generated is adjusted to $\mu=0.10$. In this case, the contribution of multi-photon events to the recorded statistics is only 4.7$\%$ and is, therefore, negligible. The generated single-photon is first sent over a single-core fiber of a DMUX, which is a device composed of $N$ single-core fibers connected to a single $N$-core fiber, where each single-core fiber is mapped to one of the cores of the MCF. The single-photon is then transmitted to one of the cores of the 4C-MCF at the end of the DMUX.

This first DMUX is connected to a 4$\times$4 MBS (4C-MBS in Fig. \ref{Figure1}). The 4$\times$4 MBSs are manufactured by locally heating a small transverse region of a homogeneous $4$-core fiber and applying a controlled longitudinal stretching tension. This leads to a tapered fiber where, due to evanescent coupling, photons will jump from one core to the others \cite{Carine}. The 4C-MBS has a mean fidelity of $F=0.995\pm0.003$ \cite{Carine} with respect to the unitary matrix 
\begin{equation}
    U_4 = \frac{1}{2} \begin{bmatrix}
    1 & 1 & 1 & 1 \\
    1 & 1 & -1 & -1 \\
    1 & -1 & 1 & -1 \\
    1 & -1 & -1 & 1
    \end{bmatrix},\label{eq:4c}
\end{equation}
while achieving an almost ideal split ratio of 25$\%$ (0.248 ± 0.01 \cite{Carine}). Thus, after the propagation through the MBS, the photon is found in a equally weighted coherent superposition of the $|i\rangle$ basis states. Next, the 4$\times$4 MBS is connected to another DMUX in reverse. Each of the 4 single-core fibers coming from this DMUX is supplemented with phase and intensity modulators (PMs and IMs, respectively) to control state parameters. While the IM voltages can be set manually, the PM voltages must be constantly adjusted as the optical paths on the single-mode fibers slowly drift. This is accomplished by having FPGA controlling the PMs with a feed-back algorithm based on counts recorded by the single-photon detectors \cite{Melo}. At this stage of the setup, single-photon states $|\chi\rangle=\sum_{i=0}^{3}\alpha_je^{i\phi_i}|i\rangle$ can be generated, thus defining the preparation stage for four-dimensional path-encoded states.

The single photons are then sent to the measurement stage through 4 single-core fibers connected to the third and different DMUX, which now maps 7 single-core fibers into a single 7-core fiber. Four of the DMUX's inputs are connected to these fibers, while the others three are not connected to any light source. The DMUX's output is connected to a 7$\times$7 MBS (7C-MBS in Fig. \ref{Figure1})) that is followed by another DMUX, which is identical to the third DMUX but in reverse. Finally, the light coming out of each of the 7 single-core fibers is detected with InGaAs single-photon detection modules (D1 through D7). The FPGA records the counts of all detectors. This combination of DMUXs, MBS, and detectors defines the measurement stage, which realizes a POVM with 7 elements acting on the 4-dimensional state $|\chi\rangle$. The elements of the POVM are dependent on which of the input fibers are used, as well as the  7$\times$7 matrix describing the MBS. The process tomography of the 7$\times$7 MBS used in our experiment was presented in Ref.~\cite{Carine} and its matrix is given explicitly in the supplemental material \cite{SuppMat}. The chosen POVM, to study experimentally the method of point tomography, is obtained when the 4 outputs of the  preparation stage are connected to the inputs $i=4,5,6$ and 7 of the measurement stage \cite{SuppMat}.

\begin{figure*}[t!]
	\centering
    	\includegraphics[width=0.9\textwidth]{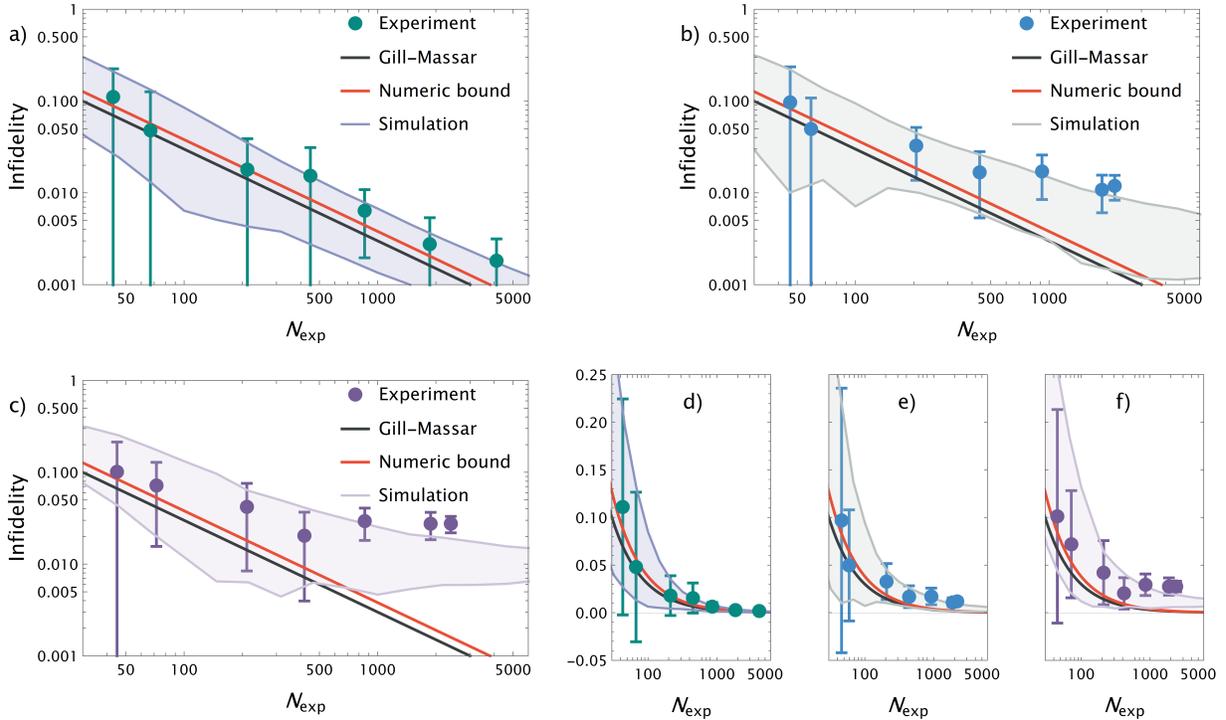}
    	\caption{Experimental results. Insets a), b), and c) are log-log plots for states $|\psi_1\rangle, |\psi_2\rangle$, and $|\psi_3\rangle$ respectively, showing the infidelity as a function of sample size. The filled circles indicate the experimentally obtained infidelity values, while error bars indicate the highest and lowest infidelity values obtained using the bootstrapping technique. The black solid line corresponds to the Gill-Massar limit for the infidelity given by $3/N$, and the red solid line is the best achievable infidelity for our setup considering a white noise error model. The shaded area shows the interquartile range generated by our error model. Insets d), e), and f) show the same data as a), b), and c), respectively, using a linear-log scale in order to show the typical reduction of the errors due to the increase in sample size.}
    	\label{Figure2}
\end{figure*}

\paragraph{\textit{Results}--} We experimentally study the estimation accuracy achievable by the method of point tomography while considering three different states $|\psi_i\rangle$ (see Eq.~(\ref{state})), with $\theta_1=10^{-2}$, $\theta_2=10^{-1}$, and $\theta_3=2\times10^{-1}$. As $\theta_i$ increases, the states $|\psi_i\rangle$ move away from the fiducial state $|0\rangle$ and the efficiency of the estimation process is expected to decrease.

The statistics generated by the measurement, for each state and for each ensemble size, is post-processed by the maximum likelihood estimation technique \cite{Shang2017}, which delivers the final estimate $|\tilde\psi_i\rangle$ of $|\psi_i\rangle$ and the corresponding infidelity $I_F=1-F(|\tilde\psi_i\rangle\langle\tilde\psi_i|,\rho_i)$, where $F(:)$ is the fidelity between quantum states. The error of the infidelity value experimentally recorded is calculated using the bootstrapping technique. To model the experimental results, we assume the traditional white noise model affecting the preparation and measurement stages. This leads to the consideration of a highly pure albeit mixed state of the form $\rho_i=\lambda|\psi_i\rangle\langle\psi_i|+(\lambda-1)I/d$  (with $\lambda=0.987$) to proper describe our experiment. We also account for systematic errors in our error model. 

The experimental results are presented in Fig.~\ref{Figure2}a, Fig.~\ref{Figure2}b, and Fig.~\ref{Figure2}c for states $|\psi_1\rangle, |\psi_2\rangle$, and $|\psi_3\rangle$, respectively, which show the infidelity as a function of the ensemble size $N$ (both axes in logarithmic scale). The filled circles indicate the experimentally obtained infidelity values, while vertical bars indicate the highest and lowest infidelity values obtained using the bootstrapping technique. The black solid line corresponds to the Gill-Massar limit for the infidelity given by $3/N$, and the red solid line is the best achievable infidelity for our setup considering the white noise error model. The shaded area shows the interquartile range generated by our error model.

Figure~\ref{Figure2}a shows the case where we estimate state $|\psi_1\rangle$, which has the largest fidelity with respect to the fiducial state $|0\rangle$.  The infidelity exhibits values very close to both the Gill-Massar bound and those of the white noise model in the inspected interval of ensemble size. This shows that the implemented POVM is able to estimate the state $|\psi_1\rangle$ with high precision following a trend similar to the Gill-Massar limit. Specifically, a fit of the experimental results related with state $|\psi_1\rangle$ leads to a infidelity behaviour that decreases as a function of the ensemble size $N$ according to $3.8/N$, which is quite close to the theoretical Gill-Massar limit given by $3/N$. Even the highest infidelity values for smaller ensembles do not differ significantly from this bound. Figures ~\ref{Figure2}b and \ref{Figure2}c show results for larger imposed systemic error. In these two cases and for small ensemble sizes, the experimental setup provides infidelity values still close to the Gill-Massar limit, and these even show a similar linear trend. As the size of the ensemble increases, the infidelity values are higher than in the case of state $|\psi_1\rangle$, and clearly differ from the lower Gill-Massar bound. Moreover, for the largest ensemble sizes, the infidelity values start to plateau. This is an experimental indication of the limits of the neighborhood that may be accurately considered in point tomography, as well as, an indication that systematic errors dominate over the effects of finite statistics for distant states. Lastly, note that the  infidelities for the estimation of $|\psi_2\rangle$ and $|\psi_3\rangle$ follow the general behavior of the white noise model (grey shaded region). Figures~\ref{Figure2}d, \ref{Figure2}e and \ref{Figure2}f are used only to show the error bars out of the logarithmic scale for each case considered, where one can see the typical reduction of the errors due to the increase in sample size.

\paragraph{\textit{Conclusion}--} We have successfully demonstrated the viability of point tomography, a state estimation technique that has practical relevance for modern high-precision quantum information processing. Specifically, assuming that an arbitrary target state can be prepared with high-accuracy by the experimentalist, the method offers a estimation efficiency that saturates the Gill-Massar limit, while requiring the implementation of a simpler POVM whose number of outcomes can be greatly reduced from the traditional value of $\sim 4d-3$ to only $2d-1$, making POVM-based tomography for higher-dimensions ($d \geq 2$) much more viable. In our work, we use state-of-the-art multicore optical fiber technology to efficiently generate four-dimensional quantum states and to implement measurements in the form of a seven-outcome generalized measurement. Our experimental investigation clearly demonstrates the viability of point tomography under real world conditions, as we studied the method in different scenarios. In the high-precision regime, we observe a precision trend close to the Gill-Massar limit even though the POVM performed is not exactly a Fisher symmetric measurement. In lower-precision regimes, we show that the method can still be relevant. For small ensemble sizes where statistical errors dominate, the average infidelities are still close to the Gill-Massar bound and show a linear decrease in the log-log plot as the ensemble size increases. The robustness of the method demonstrated in our experimental results, paves the way for a broader adoption of point tomography in all modern quantum platforms developed for high-precision quantum information.  

\begin{acknowledgments}
This work was supported by Fondo Nacional de Desarrollo Científico y Tecnológico (FONDECYT) Grants No. 1240746, 1231940, and 3200820, and by ANID -- Millennium Science Initiative Program -- ICN17$_-$012. L. P. was supported by the Government of Spain (Severo Ochoa CEX2019-000910-S, FUNQIP, Quantum in Spain, and European Union NextGenerationEU PRTR-C17.I1), Fundació Cellex, Fundació Mir-Puig, and Generalitat de Catalunya (CERCA program). KS was supported by the UCO 1866. MM was supported by ANID-PFCHA/DOCTORADO-NACIONAL/2019-21190958.
\end{acknowledgments}

\bibliography{bibpaper}% Produces the bibliography via BibTeX.

\newpage
\onecolumngrid
\section{Supplemental Material}

\subsection{Fisher-Symmetric measurements}

Here, we present the Fisher-symmetric measurements (FSM) introduced in Ref.~\cite{Li}. These measurements have been recently introduced on quantum estimation theory, which studies how much information can be obtained from observing a quantum system \cite{Helstrom1969, Belavkin1976, Yuen1973, Paris2008, Holevo2011, Liu2020}. Specifically, the goal is to determine a quantum state $\rho_\theta$, characterized by a parameter $\theta$, that will be estimated by an estimator $t$. To consider complex parameters $\theta\in \mathbb{C}^d$ without using their standard real representation, we use the complex formulation of the estimation theory \cite{Munoz2022}. It can be used for complex estimator $t$ with dependence on complex parameters $\theta$, by studying the auxiliary variables $\hat{\theta} = [\theta,\theta^*]$ and $\hat{t} = [t(\hat{\theta}), t(\hat{\theta})^*]$. One of the most remarkable results of quantum estimation theory is the quantum Cramér-Rao Inequality (QCRI). This inequality delivers a bound for the covariance matrix of an unbiased estimator implemented by a Positive Operator valued measurement (POVM) $\boldsymbol{\Pi}=\{\Pi_\omega\}$, with outcomes $\omega\in\Omega$
\begin{equation}\label{eq: QCRB}
	{\rm Cov}_{\hat{\theta}}(\boldsymbol{\Pi},\hat{t}) \geq I_{\hat\theta}^{-1} \geq J_{\hat\theta}^{-1}.
\end{equation}
Here $I_{\hat\theta}$ is the Classical Fisher Information matrix (CFIM), defined by 
\label{CCFIM}
\begin{eqnarray}
	I_{\hat\theta}=\left[\begin{array}{cc}\mathtt{I}_\theta&\mathtt{P}_\theta\\\mathtt{P}_\theta^*&\mathtt{I}_\theta^*
	\end{array}\right],
\end{eqnarray}
\begin{eqnarray}
	\left[\mathtt{I}_\theta\right]_{jk} = \sum_{\omega\in\Omega}\partial_{\theta_j^*}\ln f(\omega|\theta)\thinspace\partial_{\theta_k}\ln f(\omega|\theta), \qquad
	\left[\mathtt{P}_\theta\right]_{jk} =  \sum_{\omega\in\Omega}\partial_{\theta_j^*}\ln f(\omega|\theta)\thinspace\partial_{\theta_k^*}\ln f(\omega|\theta), 
\end{eqnarray}
where $f(\omega|\theta)=\tr( \rho_\theta \Pi_\omega )$ is the probability density function, and $J_{\hat\theta}$ is the Quantum Fisher Information Matrix (QFIM) given by 
\begin{eqnarray}\label{CSQFIM}
	J_{\hat\theta}=\left[\begin{array}{cc}\mathtt{J}_\theta & \mathtt{Q}_\theta\\
		(\mathtt{Q}_\theta)^* & (\mathtt{J}_\theta)^*\end{array}\right],
\end{eqnarray}
where the matrices $\mathtt{J}_\theta$ and $\mathtt{Q}_\theta$ for the particular case of a pure state $\rho_{\hat\theta}=\ket{\psi_{\hat\theta}}\bra{\psi_{\hat\theta}}$ are 
\begin{align}\label{CFM1_1}
	\begin{split}
	\left[\mathtt{J}_{\theta}\right]_{jk}
		=& 2\big[ \bra{\partial_{\theta^*_j}\psi_{\hat\theta}}\left(\mathbb{I}-\dyad{\psi_{\hat\theta}}{\psi_{\hat\theta}}\right)\ket{\partial_{\theta_k}\psi_{\hat\theta}}\\
		& + \bra{\partial_{\theta_k}\psi_{\hat\theta}}\left(\mathbb{I}-\dyad{\psi_{\hat\theta}}{\psi_{\hat\theta}}\right)\ket{\partial_{\theta^*_j}\psi_{\hat\theta}}  \big],	
	\end{split}
	\qquad
	\begin{split}
	[\mathtt{Q}_{\theta}]_{jk} =& 2\big[ \bra{\partial_{\theta^*_j }\psi_{\hat\theta}}\left(\mathbb{I}-\dyad{\psi_{\hat\theta}}{\psi_{\hat\theta}}\right)\ket{\partial_{\theta^*_k}\psi_{\hat\theta}}\\
	& +\bra{\partial_{\theta^*_k}\psi_{\hat\theta}}\left(\mathbb{I}-\dyad{\psi_{\hat\theta}}{\psi_{\hat\theta}}\right)\ket{\partial_{\theta^*_j}\psi_{\hat\theta}}  \big].
	\end{split}
\end{align}
For a general density matrix, these matrices depend on the Symmetric Logarithmic Derivatives (see \cite{Munoz2022}).

The Gill-Massar inequality presents the ultimate precision for the estimation of multiple parameters by individual measurements. This inequality relates the CFIM ${I}_{\hat\theta}$ and the QFIM ${J}^S_{\hat\theta}$, establishing an additional constraint to the parameter estimation problem for quantum states. It reads
\begin{equation}\label{Eq: GM Inequality}
	\tr\left({I}_{\hat\theta} {J}_{\hat\theta}^{-1}\right) \leq d-1. 
\end{equation}
Under the assumption that $\hat{t}$ attains the Classical Cramer-Rao bound ${\rm Cov}_{\hat{\theta}}(\boldsymbol{\Pi},\hat{t})={I}_{\hat\theta}^{-1}$, and using the Gill-Massar inequality as a constraint, the minimum value for the weighted mean square error (WMSE) $w_{\hat\theta}\big(\hat{t}\big) = \tr\left({W}_{\hat{\theta}}{\rm Cov}_{\hat\theta}\left(\boldsymbol{\Pi},\hat{t}\right) \right)/N_{exp}$ can be computed, where $N_{exp}$ is the sample size used to estimate the state. The optimal value is given by 
\begin{equation}
	w^{GM}=\frac{1}{N_{exp}}\frac{1}{d-1}\left(\tr\sqrt{{J}_{\hat\theta}^{-1/2}{W}_{\hat\theta}{J}_{\hat\theta}^{-1/2}} \right)^2,
\end{equation}
and is attained if and only if the measurement $\boldsymbol{\Pi}$ has that
\begin{equation}\label{eq: GM FIM}
	{I}_{\hat\theta}^{GM} = (d-1) {J}_{\hat\theta}^{1/2} \frac{\sqrt{ {J}_{\hat\theta}^{-1/2} {W}_{\hat\theta} {J}_{\hat\theta}^{-1/2}}}{{\rm Tr}\sqrt{ {J}_{\hat\theta}^{-1/2} {W}_{\hat\theta} {J}_{\hat\theta}^{-1/2}}} {J}_{\hat\theta}^{1/2}. 
\end{equation}
The WMSE agrees with the infidelity between two infinitesimally near states when ${W}_{\hat\theta}={J}_{\hat\theta}/4$. In this case, the optimal CFIM and the Gill-Massar bound for a pure state with $d-1$ complex parameters are 
\begin{align}
	w_{\hat\theta}^{GM} = \frac{d-1}{N_{exp}},\qquad {I}_{\hat\theta}^{GM} = \frac{1}{2}{J}_{\hat\theta}. \label{eq:GMB}
\end{align}

The aim of the Fisher-symmetric Measurements is to estimate a $d$-dimensional pure state $|\psi\rangle$ near a known fiducial state $|0\rangle$ with the ultimate precision provided by the Gill-Massar bound for infidelity Eq.\thinspace\eqref{eq:GMB}, but maintain a CFIM uniformly distributed over all parameters in the neighborhood of the fiducial state. Thereby, in order to obtain the FSM for $\ket{\psi}$, we first have to calculate its QFIM and CFIM. The state $|\psi\rangle$ can be parametrized up to first order as
\begin{equation}
	%\ket{\psi(\theta)}=\ket{0}+\sum_{j=1}^{d-1}\theta_j\ket{j}
	|\psi\rangle=\frac{1}{N}\left(|0\rangle+\sum_{j=1}^{3}\theta_j|j\rangle\right),
	\label{eq:psifsm}
\end{equation}
where $N\approx 1$ is a normalization constant, $\theta=[\theta_1,\thinspace \dots\thinspace, \theta_{d-1}]\in\mathbb{C}^{d-1}$ are infinitesimal complex parameters $|\theta_j|\ll 1$, and $\{\ket{j} \}$ is an orthonormal basis that include the fiducial state $\ket{0}$. The QFIMs up to first order reads
\begin{equation}
	\mathtt{J}_{jk} = 2\delta_{jk},\quad \mathtt{Q}_{jk} = 0,
\end{equation}
or equivalently ${J}=2\mathbb{I}_{2(d-1)}$. Notice that we are omitting the dependence of $\theta$ or $\hat\theta$ in the information matrices for simplicity. Let $\{E^\xi\}$ be an arbitrary Positive Operator Value Measurement (POVM) with $n$ rank one element $E^\xi=\ket{\psi^\xi}\bra{\psi^\xi}$. Considering that
\begin{equation}
	\ket{\psi^\xi}=\sum_{j=0}^{d-1}a_j^{\xi}\ket{j},\qquad \xi=0,\dots,n-1,
\end{equation}
where $a_j^{\xi}\in\mathbb{C}$ for $j\geq 1$ and $a_0^{\xi}\in\mathbb{R}$ non-negative such as $\sum_{\xi=0}^{n-1} \left( a_j^\xi \right)^* a_k^\xi = \delta_{jk}$, we have that the CFIM up to first order is given by
\begin{equation}
	\mathtt{I}_{jk}=\delta_{jk},\qquad \mathtt{P}_{jk}=\sum_{\xi=0}^{n-1}\big( a_j^\xi a_k^\xi\big)^*.
\end{equation}
The POVM $\{E^\xi\}$ defines a FSM if it satisfies the Gill-Massar bound for the infidelity ${I}={J}/2$ and has a uniform CFIM ${I}=\mathbb{I}_{2(d-1)}$. Therefore, the following conditions must be fulfilled,
\begin{equation}
	\mathtt{I}=\frac{1}{2}\mathtt{J}=\mathbb{I}_{d-1},\qquad \mathtt{P}=\frac{1}{2}\mathtt{Q}=0. 
\end{equation}
The first equation is identically satisfied, while from the second one, we have the optimality condition $\mathtt{P}=0$, which can also be written as 
\begin{equation}\label{POVMCond}
	\sum_{\xi=0}^{n-1} a_j^\xi a_k^\xi=0,\qquad j,k\geq 1.
\end{equation}
Therefore, the POVM $\{E^\xi\}$ defines a FSM if
\begin{equation}
	C_{jk}=\sum_{\xi=0}^{n-1} a_j^\xi a_k^\xi = 0.
\end{equation}

\newpage

\subsection{$7\times 7$ multiport beam splitter unitary operation}
To implement the fisher symmetric measurement required in the method of point tomography, we resort to using a $7\times 7$ multiport beam splitter (MBS) built-in seven-core multicore fibers (7C-MCFs), as explained in the main text. We also use demultiplexer/multiplexer units (DMUX) for accessing each of the seven cores of the $7\times 7$ MBS independently. The optical characterization of the $7\times 7$ MBS was done through quantum process tomography and it is presented in Ref.~\cite{Carine}. In the logical basis, the unitary matrix $\hat{U}_7$ for the MBS is given by
{\footnotesize
	\begin{equation}
	\hat{U}_7 =
	\begin{bmatrix}
	0.5639 &   0.2010 &   0.3019 &   0.3749 &   0.4918 &   0.0905 &  0.3998 \\
	0.2222 &  -0.0065 + 0.1874i&  -0.5700 - 0.3060i&   0.3558 - 0.0865i&  -0.1447 + 0.3632i&   0.2989 - 0.2884i&  -0.1033 - 0.1635i\\
	0.3487 &  -0.6271 - 0.3102i&   0.1178 - 0.0994i&  -0.2245 - 0.2686i&  -0.0469 + 0.1075i&   0.0629 - 0.2445i&  -0.0116 + 0.4061i\\
	0.3929 &   0.3320 + 0.0156i&  -0.1620 - 0.2950i&  -0.1267 + 0.3353i&  -0.0489 - 0.3414i&  -0.3319 - 0.1445i&  -0.3447 + 0.3530i\\
	0.3709 &  -0.1842 + 0.2868i&  -0.1199 + 0.1069i&  -0.0224 - 0.2699i&   0.0214 - 0.0533i&  -0.0144 + 0.7223i&  -0.3419 - 0.0698i\\
	0.1468 &   0.3709 + 0.2029i&   0.3572 - 0.0915i&  -0.0936 - 0.4318i&  -0.5262 + 0.3039i&  -0.2790 - 0.0553i&   0.1351 + 0.0108i\\
	0.4444 &  -0.0220 - 0.1704i&  -0.0651 + 0.4328i&  -0.3159 + 0.3254i&  -0.3157 - 0.0201i&   0.0839 - 0.1206i&   0.0989 - 0.4943i
	\end{bmatrix}.\nonumber \label{hatU7}
	\end{equation}
}
\noindent Here, $\hat{U}_7$ is written considering the logical basis $\{|i\rangle\}$, where $i=1,\ldots,7$ denotes the $i$-th core.

\subsection{Full list of the feasible 35 types of POVMs}

	Taking into account the unitary operation $\hat{U}_7$, our experimental setup can implement finite sets of inequivalent rank-1 POVM elements, as explained in the main text. There are 35 different families of POVMs that we can implement while four dimensional states ($d=4$) are considered. For point quantum tomography, it is necessary to find the matrix $\hat{M}_{k_i,...,K_d}$ that provides a POVM [given by Eq.(2) in the main text] whose corresponding $C$ matrix has the smallest norm. In order to check that, we minimize $C$ for each family of POVMs. After that, we identify
	{\footnotesize
		\begin{equation}
				\hat{M}_{4567}= \begin{bmatrix} 0.3749 & 0.3662 & 0.3501 & 0.3584 & 0.2709 & 0.4418 & 0.4535\\ 0.4918 & -0.2263 - 0.3188i & -0.05239 + 0.1049i & -0.3021 - 0.1664i & 0.05134 - 0.02569i & -0.1856 + 0.5787i & 0.2055 - 0.2405i\\ 0.09054 & 0.3586 + 0.2097i & 0.1473 - 0.2051i & -0.01789 - 0.3615i & -0.7187 + 0.07419i & 0.1132 + 0.261i & -0.145 - 0.02382i\\ 0.3998 & -0.06171 + 0.1833i & -0.3041 + 0.2693i & 0.4521 - 0.1977i & 0.09788 + 0.335i & -0.03915 - 0.1297i & -0.4235 - 0.2733i \end{bmatrix}, \nonumber \label{povm4567}
			\end{equation}} 
\noindent as the POVM nearest to a FSM, with $||C|| \approx 0.63$. Notice that this value is smaller than the average obtained from Haar random POVMs with 7 rank-1 elements, given by $\langle||C||\rangle\approx0.923$. That is, it is smaller than the mean value of all rank-1 POVMs generated as $|\chi_j\rangle = \sum_{j=1}^{4}U_{ij}|j\rangle$, where $U$ is a 7-dimensional Haar-random unitary matrix \cite{Mezzadri2017}. 
\bigskip

The other POVMs are:

{\footnotesize\begin{equation*}
	\hat{M}_{1234}= \begin{bmatrix} 0.5639 & 0.2222 & 0.3487 & 0.3929 & 0.3709 & 0.1468 & 0.4444\\ 0.201 & -0.006542 - 0.1874i & -0.6271 + 0.3102i & 0.332 - 0.01557i & -0.1842 - 0.2868i & 0.3709 - 0.2029i & -0.02196 + 0.1704i\\ 0.3019 & -0.57 + 0.306i & 0.1178 + 0.09943i & -0.162 + 0.295i & -0.1199 - 0.1069i & 0.3572 + 0.09154i & -0.0651 - 0.4328i\\ 0.3749 & 0.3558 + 0.08647i & -0.2245 + 0.2686i & -0.1267 - 0.3353i & -0.02242 + 0.2699i & -0.09359 + 0.4318i & -0.3159 - 0.3254i \end{bmatrix}
	\end{equation*}}

{\footnotesize\begin{equation*}  \hat{M}_{1235}= \begin{bmatrix} 0.5639 & 0.2222 & 0.3487 & 0.3929 & 0.3709 & 0.1468 & 0.4444\\ 0.201 & -0.006542 - 0.1874i & -0.6271 + 0.3102i & 0.332 - 0.01557i & -0.1842 - 0.2868i & 0.3709 - 0.2029i & -0.02196 + 0.1704i\\ 0.3019 & -0.57 + 0.306i & 0.1178 + 0.09943i & -0.162 + 0.295i & -0.1199 - 0.1069i & 0.3572 + 0.09154i & -0.0651 - 0.4328i\\ 0.4918 & -0.1447 - 0.3632i & -0.04691 - 0.1075i & -0.04889 + 0.3414i & 0.02136 + 0.0533i & -0.5262 - 0.3039i & -0.3157 + 0.02007i \end{bmatrix}
	\end{equation*}}

{\footnotesize\begin{equation*}  \hat{M}_{1236}= \begin{bmatrix} 0.5639 & 0.2222 & 0.3487 & 0.3929 & 0.3709 & 0.1468 & 0.4444\\ 0.201 & -0.006542 - 0.1874i & -0.6271 + 0.3102i & 0.332 - 0.01557i & -0.1842 - 0.2868i & 0.3709 - 0.2029i & -0.02196 + 0.1704i\\ 0.3019 & -0.57 + 0.306i & 0.1178 + 0.09943i & -0.162 + 0.295i & -0.1199 - 0.1069i & 0.3572 + 0.09154i & -0.0651 - 0.4328i\\ 0.09054 & 0.2989 + 0.2884i & 0.06289 + 0.2445i & -0.3319 + 0.1445i & -0.01444 - 0.7223i & -0.279 + 0.05535i & 0.08389 + 0.1206i \end{bmatrix}
	\end{equation*} }

{\footnotesize\begin{equation*}  \hat{M}_{1237}= \begin{bmatrix} 0.5639 & 0.2222 & 0.3487 & 0.3929 & 0.3709 & 0.1468 & 0.4444\\ 0.201 & -0.006542 - 0.1874i & -0.6271 + 0.3102i & 0.332 - 0.01557i & -0.1842 - 0.2868i & 0.3709 - 0.2029i & -0.02196 + 0.1704i\\ 0.3019 & -0.57 + 0.306i & 0.1178 + 0.09943i & -0.162 + 0.295i & -0.1199 - 0.1069i & 0.3572 + 0.09154i & -0.0651 - 0.4328i\\ 0.3998 & -0.1033 + 0.1635i & -0.01155 - 0.4061i & -0.3447 - 0.353i & -0.3419 + 0.06981i & 0.1351 - 0.01077i & 0.09887 + 0.4943i \end{bmatrix}
	\end{equation*} }

{\footnotesize\begin{equation*}  \hat{M}_{1245}= \begin{bmatrix} 0.5639 & 0.2222 & 0.3487 & 0.3929 & 0.3709 & 0.1468 & 0.4444\\ 0.201 & -0.006542 - 0.1874i & -0.6271 + 0.3102i & 0.332 - 0.01557i & -0.1842 - 0.2868i & 0.3709 - 0.2029i & -0.02196 + 0.1704i\\ 0.3749 & 0.3558 + 0.08647i & -0.2245 + 0.2686i & -0.1267 - 0.3353i & -0.02242 + 0.2699i & -0.09359 + 0.4318i & -0.3159 - 0.3254i\\ 0.4918 & -0.1447 - 0.3632i & -0.04691 - 0.1075i & -0.04889 + 0.3414i & 0.02136 + 0.0533i & -0.5262 - 0.3039i & -0.3157 + 0.02007i \end{bmatrix}
	\end{equation*} }

{\footnotesize\begin{equation*}  \hat{M}_{1246}= \begin{bmatrix} 0.5639 & 0.2222 & 0.3487 & 0.3929 & 0.3709 & 0.1468 & 0.4444\\ 0.201 & -0.006542 - 0.1874i & -0.6271 + 0.3102i & 0.332 - 0.01557i & -0.1842 - 0.2868i & 0.3709 - 0.2029i & -0.02196 + 0.1704i\\ 0.3749 & 0.3558 + 0.08647i & -0.2245 + 0.2686i & -0.1267 - 0.3353i & -0.02242 + 0.2699i & -0.09359 + 0.4318i & -0.3159 - 0.3254i\\ 0.09054 & 0.2989 + 0.2884i & 0.06289 + 0.2445i & -0.3319 + 0.1445i & -0.01444 - 0.7223i & -0.279 + 0.05535i & 0.08389 + 0.1206i \end{bmatrix}
	\end{equation*} }

{\footnotesize\begin{equation*}  \hat{M}_{1247}= \begin{bmatrix} 0.5639 & 0.2222 & 0.3487 & 0.3929 & 0.3709 & 0.1468 & 0.4444\\ 0.201 & -0.006542 - 0.1874i & -0.6271 + 0.3102i & 0.332 - 0.01557i & -0.1842 - 0.2868i & 0.3709 - 0.2029i & -0.02196 + 0.1704i\\ 0.3749 & 0.3558 + 0.08647i & -0.2245 + 0.2686i & -0.1267 - 0.3353i & -0.02242 + 0.2699i & -0.09359 + 0.4318i & -0.3159 - 0.3254i\\ 0.3998 & -0.1033 + 0.1635i & -0.01155 - 0.4061i & -0.3447 - 0.353i & -0.3419 + 0.06981i & 0.1351 - 0.01077i & 0.09887 + 0.4943i \end{bmatrix}
	\end{equation*} }

{\footnotesize\begin{equation*}  \hat{M}_{1256}= \begin{bmatrix} 0.5639 & 0.2222 & 0.3487 & 0.3929 & 0.3709 & 0.1468 & 0.4444\\ 0.201 & -0.006542 - 0.1874i & -0.6271 + 0.3102i & 0.332 - 0.01557i & -0.1842 - 0.2868i & 0.3709 - 0.2029i & -0.02196 + 0.1704i\\ 0.4918 & -0.1447 - 0.3632i & -0.04691 - 0.1075i & -0.04889 + 0.3414i & 0.02136 + 0.0533i & -0.5262 - 0.3039i & -0.3157 + 0.02007i\\ 0.09054 & 0.2989 + 0.2884i & 0.06289 + 0.2445i & -0.3319 + 0.1445i & -0.01444 - 0.7223i & -0.279 + 0.05535i & 0.08389 + 0.1206i \end{bmatrix}
	\end{equation*} }

{\footnotesize\begin{equation*}  \hat{M}_{1257}= \begin{bmatrix} 0.5639 & 0.2222 & 0.3487 & 0.3929 & 0.3709 & 0.1468 & 0.4444\\ 0.201 & -0.006542 - 0.1874i & -0.6271 + 0.3102i & 0.332 - 0.01557i & -0.1842 - 0.2868i & 0.3709 - 0.2029i & -0.02196 + 0.1704i\\ 0.4918 & -0.1447 - 0.3632i & -0.04691 - 0.1075i & -0.04889 + 0.3414i & 0.02136 + 0.0533i & -0.5262 - 0.3039i & -0.3157 + 0.02007i\\ 0.3998 & -0.1033 + 0.1635i & -0.01155 - 0.4061i & -0.3447 - 0.353i & -0.3419 + 0.06981i & 0.1351 - 0.01077i & 0.09887 + 0.4943i \end{bmatrix}
	\end{equation*} }

{\footnotesize\begin{equation*}  \hat{M}_{1267}= \begin{bmatrix} 0.5639 & 0.2222 & 0.3487 & 0.3929 & 0.3709 & 0.1468 & 0.4444\\ 0.201 & -0.006542 - 0.1874i & -0.6271 + 0.3102i & 0.332 - 0.01557i & -0.1842 - 0.2868i & 0.3709 - 0.2029i & -0.02196 + 0.1704i\\ 0.09054 & 0.2989 + 0.2884i & 0.06289 + 0.2445i & -0.3319 + 0.1445i & -0.01444 - 0.7223i & -0.279 + 0.05535i & 0.08389 + 0.1206i\\ 0.3998 & -0.1033 + 0.1635i & -0.01155 - 0.4061i & -0.3447 - 0.353i & -0.3419 + 0.06981i & 0.1351 - 0.01077i & 0.09887 + 0.4943i \end{bmatrix}
	\end{equation*} }

{\footnotesize\begin{equation*}  \hat{M}_{1345}= \begin{bmatrix} 0.5639 & 0.2222 & 0.3487 & 0.3929 & 0.3709 & 0.1468 & 0.4444\\ 0.3019 & -0.57 + 0.306i & 0.1178 + 0.09943i & -0.162 + 0.295i & -0.1199 - 0.1069i & 0.3572 + 0.09154i & -0.0651 - 0.4328i\\ 0.3749 & 0.3558 + 0.08647i & -0.2245 + 0.2686i & -0.1267 - 0.3353i & -0.02242 + 0.2699i & -0.09359 + 0.4318i & -0.3159 - 0.3254i\\ 0.4918 & -0.1447 - 0.3632i & -0.04691 - 0.1075i & -0.04889 + 0.3414i & 0.02136 + 0.0533i & -0.5262 - 0.3039i & -0.3157 + 0.02007i \end{bmatrix}
	\end{equation*} }

{\footnotesize\begin{equation*}  \hat{M}_{1346}= \begin{bmatrix} 0.5639 & 0.2222 & 0.3487 & 0.3929 & 0.3709 & 0.1468 & 0.4444\\ 0.3019 & -0.57 + 0.306i & 0.1178 + 0.09943i & -0.162 + 0.295i & -0.1199 - 0.1069i & 0.3572 + 0.09154i & -0.0651 - 0.4328i\\ 0.3749 & 0.3558 + 0.08647i & -0.2245 + 0.2686i & -0.1267 - 0.3353i & -0.02242 + 0.2699i & -0.09359 + 0.4318i & -0.3159 - 0.3254i\\ 0.09054 & 0.2989 + 0.2884i & 0.06289 + 0.2445i & -0.3319 + 0.1445i & -0.01444 - 0.7223i & -0.279 + 0.05535i & 0.08389 + 0.1206i \end{bmatrix}
	\end{equation*} }

{\footnotesize\begin{equation*}  \hat{M}_{1347}= \begin{bmatrix} 0.5639 & 0.2222 & 0.3487 & 0.3929 & 0.3709 & 0.1468 & 0.4444\\ 0.3019 & -0.57 + 0.306i & 0.1178 + 0.09943i & -0.162 + 0.295i & -0.1199 - 0.1069i & 0.3572 + 0.09154i & -0.0651 - 0.4328i\\ 0.3749 & 0.3558 + 0.08647i & -0.2245 + 0.2686i & -0.1267 - 0.3353i & -0.02242 + 0.2699i & -0.09359 + 0.4318i & -0.3159 - 0.3254i\\ 0.3998 & -0.1033 + 0.1635i & -0.01155 - 0.4061i & -0.3447 - 0.353i & -0.3419 + 0.06981i & 0.1351 - 0.01077i & 0.09887 + 0.4943i \end{bmatrix}
	\end{equation*} }

{\footnotesize\begin{equation*}  \hat{M}_{1356}= \begin{bmatrix} 0.5639 & 0.2222 & 0.3487 & 0.3929 & 0.3709 & 0.1468 & 0.4444\\ 0.3019 & -0.57 + 0.306i & 0.1178 + 0.09943i & -0.162 + 0.295i & -0.1199 - 0.1069i & 0.3572 + 0.09154i & -0.0651 - 0.4328i\\ 0.4918 & -0.1447 - 0.3632i & -0.04691 - 0.1075i & -0.04889 + 0.3414i & 0.02136 + 0.0533i & -0.5262 - 0.3039i & -0.3157 + 0.02007i\\ 0.09054 & 0.2989 + 0.2884i & 0.06289 + 0.2445i & -0.3319 + 0.1445i & -0.01444 - 0.7223i & -0.279 + 0.05535i & 0.08389 + 0.1206i \end{bmatrix}
	\end{equation*} }

{\footnotesize\begin{equation*}  \hat{M}_{1357}= \begin{bmatrix} 0.5639 & 0.2222 & 0.3487 & 0.3929 & 0.3709 & 0.1468 & 0.4444\\ 0.3019 & -0.57 + 0.306i & 0.1178 + 0.09943i & -0.162 + 0.295i & -0.1199 - 0.1069i & 0.3572 + 0.09154i & -0.0651 - 0.4328i\\ 0.4918 & -0.1447 - 0.3632i & -0.04691 - 0.1075i & -0.04889 + 0.3414i & 0.02136 + 0.0533i & -0.5262 - 0.3039i & -0.3157 + 0.02007i\\ 0.3998 & -0.1033 + 0.1635i & -0.01155 - 0.4061i & -0.3447 - 0.353i & -0.3419 + 0.06981i & 0.1351 - 0.01077i & 0.09887 + 0.4943i \end{bmatrix}
	\end{equation*} }

{\footnotesize\begin{equation*}  \hat{M}_{1367}= \begin{bmatrix} 0.5639 & 0.2222 & 0.3487 & 0.3929 & 0.3709 & 0.1468 & 0.4444\\ 0.3019 & -0.57 + 0.306i & 0.1178 + 0.09943i & -0.162 + 0.295i & -0.1199 - 0.1069i & 0.3572 + 0.09154i & -0.0651 - 0.4328i\\ 0.09054 & 0.2989 + 0.2884i & 0.06289 + 0.2445i & -0.3319 + 0.1445i & -0.01444 - 0.7223i & -0.279 + 0.05535i & 0.08389 + 0.1206i\\ 0.3998 & -0.1033 + 0.1635i & -0.01155 - 0.4061i & -0.3447 - 0.353i & -0.3419 + 0.06981i & 0.1351 - 0.01077i & 0.09887 + 0.4943i \end{bmatrix}
	\end{equation*} }

{\footnotesize\begin{equation*}  \hat{M}_{1456}= \begin{bmatrix} 0.5639 & 0.2222 & 0.3487 & 0.3929 & 0.3709 & 0.1468 & 0.4444\\ 0.3749 & 0.3558 + 0.08647i & -0.2245 + 0.2686i & -0.1267 - 0.3353i & -0.02242 + 0.2699i & -0.09359 + 0.4318i & -0.3159 - 0.3254i\\ 0.4918 & -0.1447 - 0.3632i & -0.04691 - 0.1075i & -0.04889 + 0.3414i & 0.02136 + 0.0533i & -0.5262 - 0.3039i & -0.3157 + 0.02007i\\ 0.09054 & 0.2989 + 0.2884i & 0.06289 + 0.2445i & -0.3319 + 0.1445i & -0.01444 - 0.7223i & -0.279 + 0.05535i & 0.08389 + 0.1206i \end{bmatrix}
	\end{equation*} }

{\footnotesize\begin{equation*}  \hat{M}_{1457}= \begin{bmatrix} 0.5639 & 0.2222 & 0.3487 & 0.3929 & 0.3709 & 0.1468 & 0.4444\\ 0.3749 & 0.3558 + 0.08647i & -0.2245 + 0.2686i & -0.1267 - 0.3353i & -0.02242 + 0.2699i & -0.09359 + 0.4318i & -0.3159 - 0.3254i\\ 0.4918 & -0.1447 - 0.3632i & -0.04691 - 0.1075i & -0.04889 + 0.3414i & 0.02136 + 0.0533i & -0.5262 - 0.3039i & -0.3157 + 0.02007i\\ 0.3998 & -0.1033 + 0.1635i & -0.01155 - 0.4061i & -0.3447 - 0.353i & -0.3419 + 0.06981i & 0.1351 - 0.01077i & 0.09887 + 0.4943i \end{bmatrix}
	\end{equation*} }

{\footnotesize\begin{equation*}  \hat{M}_{1467}= \begin{bmatrix} 0.5639 & 0.2222 & 0.3487 & 0.3929 & 0.3709 & 0.1468 & 0.4444\\ 0.3749 & 0.3558 + 0.08647i & -0.2245 + 0.2686i & -0.1267 - 0.3353i & -0.02242 + 0.2699i & -0.09359 + 0.4318i & -0.3159 - 0.3254i\\ 0.09054 & 0.2989 + 0.2884i & 0.06289 + 0.2445i & -0.3319 + 0.1445i & -0.01444 - 0.7223i & -0.279 + 0.05535i & 0.08389 + 0.1206i\\ 0.3998 & -0.1033 + 0.1635i & -0.01155 - 0.4061i & -0.3447 - 0.353i & -0.3419 + 0.06981i & 0.1351 - 0.01077i & 0.09887 + 0.4943i \end{bmatrix}
	\end{equation*} }

{\footnotesize\begin{equation*}  \hat{M}_{1567}= \begin{bmatrix} 0.5639 & 0.2222 & 0.3487 & 0.3929 & 0.3709 & 0.1468 & 0.4444\\ 0.4918 & -0.1447 - 0.3632i & -0.04691 - 0.1075i & -0.04889 + 0.3414i & 0.02136 + 0.0533i & -0.5262 - 0.3039i & -0.3157 + 0.02007i\\ 0.09054 & 0.2989 + 0.2884i & 0.06289 + 0.2445i & -0.3319 + 0.1445i & -0.01444 - 0.7223i & -0.279 + 0.05535i & 0.08389 + 0.1206i\\ 0.3998 & -0.1033 + 0.1635i & -0.01155 - 0.4061i & -0.3447 - 0.353i & -0.3419 + 0.06981i & 0.1351 - 0.01077i & 0.09887 + 0.4943i \end{bmatrix}
	\end{equation*} }

{\footnotesize\begin{equation*}  \hat{M}_{2345}= \begin{bmatrix} 0.201 & 0.1875 & 0.6997 & 0.3324 & 0.3409 & 0.4227 & 0.1718\\ 0.3019 & -0.2859 - 0.5804i & -0.0615 - 0.1414i & -0.1756 + 0.2871i & 0.1548 - 0.04311i & 0.2694 + 0.2517i & -0.4209 + 0.1199i\\ 0.3749 & -0.09883 + 0.3526i & 0.3204 - 0.1412i & -0.1108 - 0.3409i & -0.215 - 0.1647i & -0.2893 + 0.3339i & -0.2824 + 0.3549i\\ 0.4918 & 0.368 - 0.1319i & -0.005624 + 0.1171i & -0.06483 + 0.3387i & -0.05638 - 0.01083i & -0.3158 - 0.5192i & 0.06025 + 0.3105i \end{bmatrix}
	\end{equation*} }

{\footnotesize\begin{equation*}  \hat{M}_{2346}= \begin{bmatrix} 0.201 & 0.1875 & 0.6997 & 0.3324 & 0.3409 & 0.4227 & 0.1718\\ 0.3019 & -0.2859 - 0.5804i & -0.0615 - 0.1414i & -0.1756 + 0.2871i & 0.1548 - 0.04311i & 0.2694 + 0.2517i & -0.4209 + 0.1199i\\ 0.3749 & -0.09883 + 0.3526i & 0.3204 - 0.1412i & -0.1108 - 0.3409i & -0.215 - 0.1647i & -0.2893 + 0.3339i & -0.2824 + 0.3549i\\ 0.09054 & -0.2987 + 0.2887i & 0.05206 - 0.2471i & -0.3383 + 0.1288i & 0.6156 + 0.3782i & -0.2714 - 0.08536i & 0.1089 - 0.09863i \end{bmatrix}
	\end{equation*} }

{\footnotesize\begin{equation*}  \hat{M}_{2347}= \begin{bmatrix} 0.201 & 0.1875 & 0.6997 & 0.3324 & 0.3409 & 0.4227 & 0.1718\\ 0.3019 & -0.2859 - 0.5804i & -0.0615 - 0.1414i & -0.1756 + 0.2871i & 0.1548 - 0.04311i & 0.2694 + 0.2517i & -0.4209 + 0.1199i\\ 0.3749 & -0.09883 + 0.3526i & 0.3204 - 0.1412i & -0.1108 - 0.3409i & -0.215 - 0.1647i & -0.2893 + 0.3339i & -0.2824 + 0.3549i\\ 0.3998 & -0.1598 - 0.1089i & -0.1697 + 0.3691i & -0.3278 - 0.3688i & 0.1261 - 0.3254i & 0.1237 + 0.05538i & 0.4776 - 0.1612i \end{bmatrix}
	\end{equation*} }

{\footnotesize\begin{equation*}  \hat{M}_{2356}= \begin{bmatrix} 0.201 & 0.1875 & 0.6997 & 0.3324 & 0.3409 & 0.4227 & 0.1718\\ 0.3019 & -0.2859 - 0.5804i & -0.0615 - 0.1414i & -0.1756 + 0.2871i & 0.1548 - 0.04311i & 0.2694 + 0.2517i & -0.4209 + 0.1199i\\ 0.4918 & 0.368 - 0.1319i & -0.005624 + 0.1171i & -0.06483 + 0.3387i & -0.05638 - 0.01083i & -0.3158 - 0.5192i & 0.06025 + 0.3105i\\ 0.09054 & -0.2987 + 0.2887i & 0.05206 - 0.2471i & -0.3383 + 0.1288i & 0.6156 + 0.3782i & -0.2714 - 0.08536i & 0.1089 - 0.09863i \end{bmatrix}
	\end{equation*} }

{\footnotesize\begin{equation*}  \hat{M}_{2357}= \begin{bmatrix} 0.201 & 0.1875 & 0.6997 & 0.3324 & 0.3409 & 0.4227 & 0.1718\\ 0.3019 & -0.2859 - 0.5804i & -0.0615 - 0.1414i & -0.1756 + 0.2871i & 0.1548 - 0.04311i & 0.2694 + 0.2517i & -0.4209 + 0.1199i\\ 0.4918 & 0.368 - 0.1319i & -0.005624 + 0.1171i & -0.06483 + 0.3387i & -0.05638 - 0.01083i & -0.3158 - 0.5192i & 0.06025 + 0.3105i\\ 0.3998 & -0.1598 - 0.1089i & -0.1697 + 0.3691i & -0.3278 - 0.3688i & 0.1261 - 0.3254i & 0.1237 + 0.05538i & 0.4776 - 0.1612i \end{bmatrix}
	\end{equation*} }

{\footnotesize\begin{equation*}  \hat{M}_{2367}= \begin{bmatrix} 0.201 & 0.1875 & 0.6997 & 0.3324 & 0.3409 & 0.4227 & 0.1718\\ 0.3019 & -0.2859 - 0.5804i & -0.0615 - 0.1414i & -0.1756 + 0.2871i & 0.1548 - 0.04311i & 0.2694 + 0.2517i & -0.4209 + 0.1199i\\ 0.09054 & -0.2987 + 0.2887i & 0.05206 - 0.2471i & -0.3383 + 0.1288i & 0.6156 + 0.3782i & -0.2714 - 0.08536i & 0.1089 - 0.09863i\\ 0.3998 & -0.1598 - 0.1089i & -0.1697 + 0.3691i & -0.3278 - 0.3688i & 0.1261 - 0.3254i & 0.1237 + 0.05538i & 0.4776 - 0.1612i \end{bmatrix}
	\end{equation*} }

{\footnotesize\begin{equation*}  \hat{M}_{2456}= \begin{bmatrix} 0.201 & 0.1875 & 0.6997 & 0.3324 & 0.3409 & 0.4227 & 0.1718\\ 0.3749 & -0.09883 + 0.3526i & 0.3204 - 0.1412i & -0.1108 - 0.3409i & -0.215 - 0.1647i & -0.2893 + 0.3339i & -0.2824 + 0.3549i\\ 0.4918 & 0.368 - 0.1319i & -0.005624 + 0.1171i & -0.06483 + 0.3387i & -0.05638 - 0.01083i & -0.3158 - 0.5192i & 0.06025 + 0.3105i\\ 0.09054 & -0.2987 + 0.2887i & 0.05206 - 0.2471i & -0.3383 + 0.1288i & 0.6156 + 0.3782i & -0.2714 - 0.08536i & 0.1089 - 0.09863i \end{bmatrix}
	\end{equation*} }

{\footnotesize\begin{equation*}  \hat{M}_{2457}= \begin{bmatrix} 0.201 & 0.1875 & 0.6997 & 0.3324 & 0.3409 & 0.4227 & 0.1718\\ 0.3749 & -0.09883 + 0.3526i & 0.3204 - 0.1412i & -0.1108 - 0.3409i & -0.215 - 0.1647i & -0.2893 + 0.3339i & -0.2824 + 0.3549i\\ 0.4918 & 0.368 - 0.1319i & -0.005624 + 0.1171i & -0.06483 + 0.3387i & -0.05638 - 0.01083i & -0.3158 - 0.5192i & 0.06025 + 0.3105i\\ 0.3998 & -0.1598 - 0.1089i & -0.1697 + 0.3691i & -0.3278 - 0.3688i & 0.1261 - 0.3254i & 0.1237 + 0.05538i & 0.4776 - 0.1612i \end{bmatrix}
	\end{equation*} }

{\footnotesize\begin{equation*}  \hat{M}_{2467}= \begin{bmatrix} 0.201 & 0.1875 & 0.6997 & 0.3324 & 0.3409 & 0.4227 & 0.1718\\ 0.3749 & -0.09883 + 0.3526i & 0.3204 - 0.1412i & -0.1108 - 0.3409i & -0.215 - 0.1647i & -0.2893 + 0.3339i & -0.2824 + 0.3549i\\ 0.09054 & -0.2987 + 0.2887i & 0.05206 - 0.2471i & -0.3383 + 0.1288i & 0.6156 + 0.3782i & -0.2714 - 0.08536i & 0.1089 - 0.09863i\\ 0.3998 & -0.1598 - 0.1089i & -0.1697 + 0.3691i & -0.3278 - 0.3688i & 0.1261 - 0.3254i & 0.1237 + 0.05538i & 0.4776 - 0.1612i \end{bmatrix}
	\end{equation*} }

{\footnotesize\begin{equation*}  \hat{M}_{2567}= \begin{bmatrix} 0.201 & 0.1875 & 0.6997 & 0.3324 & 0.3409 & 0.4227 & 0.1718\\ 0.4918 & 0.368 - 0.1319i & -0.005624 + 0.1171i & -0.06483 + 0.3387i & -0.05638 - 0.01083i & -0.3158 - 0.5192i & 0.06025 + 0.3105i\\ 0.09054 & -0.2987 + 0.2887i & 0.05206 - 0.2471i & -0.3383 + 0.1288i & 0.6156 + 0.3782i & -0.2714 - 0.08536i & 0.1089 - 0.09863i\\ 0.3998 & -0.1598 - 0.1089i & -0.1697 + 0.3691i & -0.3278 - 0.3688i & 0.1261 - 0.3254i & 0.1237 + 0.05538i & 0.4776 - 0.1612i \end{bmatrix}
	\end{equation*} }

{\footnotesize\begin{equation*}  \hat{M}_{3456}= \begin{bmatrix} 0.3019 & 0.647 & 0.1542 & 0.3365 & 0.1607 & 0.3687 & 0.4377\\ 0.3749 & -0.2726 - 0.2445i & 0.001653 + 0.3501i & -0.2329 + 0.2725i & -0.1629 - 0.2164i & 0.01653 + 0.4415i & 0.3688 - 0.264i\\ 0.4918 & -0.0443 + 0.3884i & -0.1052 - 0.05189i & 0.3228 - 0.1215i & -0.05141 - 0.02556i & -0.5852 - 0.1638i & 0.02711 - 0.3152i\\ 0.09054 & -0.1269 - 0.3955i & 0.2058 + 0.1463i & 0.2864 + 0.2213i & 0.4915 + 0.5295i & -0.2566 + 0.1229i & -0.1318 + 0.06502i \end{bmatrix}
	\end{equation*} }

{\footnotesize\begin{equation*}  \hat{M}_{3457}= \begin{bmatrix} 0.3019 & 0.647 & 0.1542 & 0.3365 & 0.1607 & 0.3687 & 0.4377\\ 0.3749 & -0.2726 - 0.2445i & 0.001653 + 0.3501i & -0.2329 + 0.2725i & -0.1629 - 0.2164i & 0.01653 + 0.4415i & 0.3688 - 0.264i\\ 0.4918 & -0.0443 + 0.3884i & -0.1052 - 0.05189i & 0.3228 - 0.1215i & -0.05141 - 0.02556i & -0.5852 - 0.1638i & 0.02711 - 0.3152i\\ 0.3998 & 0.1683 - 0.09526i & -0.2707 - 0.3029i & -0.1435 + 0.4721i & 0.2088 - 0.2797i & 0.1282 - 0.04397i & -0.5035 + 0.02425i \end{bmatrix}
	\end{equation*} }

{\footnotesize\begin{equation*}  \hat{M}_{3467}= \begin{bmatrix} 0.3019 & 0.647 & 0.1542 & 0.3365 & 0.1607 & 0.3687 & 0.4377\\ 0.3749 & -0.2726 - 0.2445i & 0.001653 + 0.3501i & -0.2329 + 0.2725i & -0.1629 - 0.2164i & 0.01653 + 0.4415i & 0.3688 - 0.264i\\ 0.09054 & -0.1269 - 0.3955i & 0.2058 + 0.1463i & 0.2864 + 0.2213i & 0.4915 + 0.5295i & -0.2566 + 0.1229i & -0.1318 + 0.06502i\\ 0.3998 & 0.1683 - 0.09526i & -0.2707 - 0.3029i & -0.1435 + 0.4721i & 0.2088 - 0.2797i & 0.1282 - 0.04397i & -0.5035 + 0.02425i \end{bmatrix}
	\end{equation*} }

{\footnotesize\begin{equation*}  \hat{M}_{3567}= \begin{bmatrix} 0.3019 & 0.647 & 0.1542 & 0.3365 & 0.1607 & 0.3687 & 0.4377\\ 0.4918 & -0.0443 + 0.3884i & -0.1052 - 0.05189i & 0.3228 - 0.1215i & -0.05141 - 0.02556i & -0.5852 - 0.1638i & 0.02711 - 0.3152i\\ 0.09054 & -0.1269 - 0.3955i & 0.2058 + 0.1463i & 0.2864 + 0.2213i & 0.4915 + 0.5295i & -0.2566 + 0.1229i & -0.1318 + 0.06502i\\ 0.3998 & 0.1683 - 0.09526i & -0.2707 - 0.3029i & -0.1435 + 0.4721i & 0.2088 - 0.2797i & 0.1282 - 0.04397i & -0.5035 + 0.02425i \end{bmatrix}
	\end{equation*} }

\end{document}